\documentstyle[12pt,epsf]{article}
\newcommand{\nn}{\nonumber}
\newcommand{\tr}{\mbox{tr}}
\newcommand{\eqn}[1]{(\ref{#1})}

\newcommand{\real}{{\bb R}} 
\newcommand{\mod}{{\bb M}} 
\newcommand{\mods}{{\bbs M}} 

\font\mybb=msbm10 at 12pt
\def\bb#1{\hbox{\mybb#1}}
\font\mybbs=msbm10 at 9pt
\def\bbs#1{\hbox{\mybbs#1}}

\def\geqsim{{>\!\!\!\!\!^{\,}_\sim}}
\def\e{{\rm e}}

\def\beq{\begin{equation}}
\def\eeq{\end{equation}}
\def\bea{\begin{eqnarray}}
\def\eea{\end{eqnarray}}
\def\bd{\begin{displaymath}}
\def\ed{\end{displaymath}}

\makeatletter
\newdimen\normalarrayskip              
\newdimen\minarrayskip                 
\normalarrayskip\baselineskip
\minarrayskip\jot
\newif\ifold             \oldtrue            
\def\arraymode{\ifold\relax\else\displaystyle\fi} 
\def\@arrayskip{\ifold\baselineskip\z@\lineskip\z@
     \else
     \baselineskip\minarrayskip\lineskip2\minarrayskip\fi}
\def\@arrayclassz{\ifcase \@lastchclass \@acolampacol \or
\@ampacol \or \or \or \@addamp \or
   \@acolampacol \or \@firstampfalse \@acol \fi
\edef\@preamble{\@preamble
  \ifcase \@chnum
     \hfil$\relax\arraymode\@sharp$\hfil
     \or $\relax\arraymode\@sharp$\hfil
     \or \hfil$\relax\arraymode\@sharp$\fi}}
\def\@array[#1]#2{\setbox\@arstrutbox=\hbox{\vrule
     height\arraystretch \ht\strutbox
     depth\arraystretch \dp\strutbox
     width\z@}\@mkpream{#2}\edef\@preamble{\halign \noexpand\@halignto
\bgroup \tabskip\z@ \@arstrut \@preamble \tabskip\z@ \cr}%
\let\@startpbox\@@startpbox \let\@endpbox\@@endpbox
  \if #1t\vtop \else \if#1b\vbox \else \vcenter \fi\fi
  \bgroup \let\par\relax
  \let\@sharp##\let\protect\relax
  \@arrayskip\@preamble}
\makeatother

\newcommand{\newsection}[1]
{\vspace{4mm}
\pagebreak[3]
\addtocounter{section}{1}
\setcounter{equation}{0}
\setcounter{subsection}{0}
\begin{flushleft}
{\large\bf \thesection. #1}
\end{flushleft}
\nopagebreak
\medskip
\nopagebreak}

\newlength{\extraspace}
\newlength{\extraspaces}
\begin{document}

\pagestyle{plain}
\setcounter{footnote}{0}
\setcounter{page}{1}
\stepcounter{subsection}

\renewcommand{\footnotesize}{\small}

\addtolength{\baselineskip}{.7mm}

\thispagestyle{empty}

\begin{flushright}
\baselineskip=12pt
OUTP-98-76P\\
NBI-HE-98-36\\
hep-th/9811116\\
\hfill{  }\\ November 1998
\end{flushright}

\vskip 0.1in

\begin{center}

\baselineskip=18pt

{\large\bf{D-BRANES AND THE NON-COMMUTATIVE STRUCTURE OF QUANTUM
SPACETIME}\footnote{\baselineskip=12ptBased on talks given by {\sc r.j.s.} at
{\it SUSY '98}, Oxford, England, July 11--17, 1998, and by {\sc n.e.m.} at the
Corfu Summer Institute: 6th Hellenic School and Workshop on Elementary Particle
Physics, TMR project {\it Physics Beyond the Standard Model}, Corfu, Greece,
September 15--18, 1998.}}\\[8mm]

\baselineskip=12pt

{\bf Nick E. Mavromatos}\footnote{\baselineskip=12pt PPARC Advanced Fellow
(U.K.).}
\\[3mm]
{\it Department of Physics -- Theoretical Physics, University of Oxford\\ 1
Keble Road, Oxford OX1 3NP, U.K.}\\ {\tt n.mavromatos1@physics.oxford.ac.uk}
\\[6mm]
{\bf Richard J.\ Szabo}
\\[3mm]
{\it The Niels Bohr Institute\\ Blegdamsvej 17, DK-2100 Copenhagen \O,
Denmark}\\ {\tt szabo@nbi.dk}
\\[10mm]

\begin{center}
\begin{minipage}{14cm}

\small

\baselineskip=12pt

A worldsheet approach to the study of non-abelian D-particle dynamics is
presented based on viewing matrix-valued D-brane coordinate fields as coupling
constants of a deformed $\sigma$-model which defines a logarithmic conformal
field theory. The short-distance structure of spacetime is shown to be
naturally captured by the Zamolodchikov metric on the corresponding moduli
space which encodes the geometry of the string interactions between
D-particles. Spacetime quantization is induced directly by the string genus
expansion and leads to new forms of uncertainty relations which imply that
general relativity at very short-distance scales is intrinsically described by
a non-commutative geometry. The indeterminancies exhibit decoherence effects
suggesting the natural incorporation of quantum gravity by short-distance
D-particle probes. Some potential experimental tests are briefly described.

\end{minipage}
\end{center}

\end{center}

\bigskip

\baselineskip=14pt

\newsection{New Uncertainty Principles in String Theory}

A long-standing problem in string theory is to determine the structure of
spacetime at very short distance scales, typically at lengths smaller than the
finite intrinsic length of the strings. One of the first analytical approaches
to this problem was to study the effects of high-energy string scattering
amplitudes on the accuracy with which one can measure position and momentum
\cite{ven}. This implies the conventional string-modified Heisenberg
uncertainty principle
\beq
\Delta x~\geqsim~\frac\hbar{\Delta p}+{\cal O}(\ell_s^2)\,\Delta p+\dots
\label{stringheisen}\eeq
The modifications to the usual phase space uncertainty relation in
\eqn{stringheisen} come from stringy corrections which are due to the finite
minimum length $\ell_s$ of the string. In fact, minimizing the right-hand side
of \eqn{stringheisen} shows that the string length scale gives an absolute
minimum lower bound $\Delta x\geq{\cal O}(\ell_s)$ on the measurability of
distances in spacetime. This result means that, if one uses only string states
as probes of short-distance structure, the conventional ideas of general
relativity break down at distances smaller than $\ell_s$.

However, until very recently there has been no systematic derivation of
\eqn{stringheisen} based on some set of fundamental principles. The appearence
of new solitonic structures in string theory, which incorporate defects in
spacetime, suggest the possibility of using such objects as probes of
short-distance structure. These non-perturbative objects are known as D-branes
and can be analytically described beyond the conventional string worldsheet
approach. In many instances however, such as the cases that will be studied in
the following, a perturbative string loop-expansion approach is still
sufficient. As we will discuss in this paper, such an approach leads to new
forms of uncertainty relations, in addition to \eqn{stringheisen}, which are
attributed to the recoil of the spacetime defects in the process of the
scattering of string matter off the D-brane solitons. In the case of
multi-brane systems, this leads to a non-commutative spacetime structure at
very short distances.

In this paper we will give an exposition of these results, based mostly on the
articles \cite{emnd}--\cite{ms}. We will begin with a brief review of soliton
structures in string theory, emphasizing the worldsheet $\sigma$-model approach
to $T$-duality and Dirichlet branes. We will then describe how to view
spacetime coordinates and momenta in this framework as $\sigma$-model coupling
constants, such that the genus expansion leads to a canonical phase space
quantization. We then specialize to the case of a system of multi-D-particles
and show how one passes from Lie algebraic to spacetime non-commutativity. In
this way the quantum spacetime which follows from the many-body D-particle
dynamics is induced directly by the quantum string theory itself. The
important aspects of the construction are a logarithmic conformal field theory
formalism for the relevant recoil operators, an effective target space
Lagrangian for the $\sigma$-model couplings which is described by non-abelian
Born-Infeld theory, and the interpretation of the target space time as the
Liouville mode of the underlying two-dimensional quantum gravity. We will show
how this formalism leads to new uncertainty principles in D-brane quantum
gravity. We conclude with some general remarks and outlook on the nature of
time in D-brane quantum gravity, including possible experimental tests of
spacetime non-commutativity from $\gamma$-ray burst observations, neutral kaon
physics, and atom interferometry measurements.

\subsubsection*{Quantization of Collective Coordinates: The Basic Idea}

It is worthwhile to give a quick overview of the formalism which will follow.
To use D-branes as probes of the short distance properties of spacetime, we
shall view the collective coordinates and momenta of D-particles as a set of
$\sigma$-model couplings $\{g^I\}$ (this includes the case of multi-D0-brane
configurations which inherently contain non-commutative structures). It follows
from the Coleman approach to probabilistic couplings via two-dimensional
quantum gravity wormholes \cite{wormholes}, that the genus expansion of the
worldsheet theory will lead to a quantization of the couplings $\{g^I\}$
\cite{emn}.

\begin{figure}[htb]
\unitlength=1.00mm
\linethickness{0.4pt}
\bigskip
\begin{center}
\begin{picture}(70.00,10.00)
\LARGE
\put(5.00,2.00){\circle*{100.00}}
\put(8.00,2.00){\circle{3.00}}
\put(15.00,2.00){\makebox(0,0)[l]{$+$}}
\put(28.00,2.00){\circle*{100.00}}
\put(31.00,2.00){\circle{3.00}}
\put(28.00,5.00){\circle{3.00}}
\put(38.00,2.00){\makebox(0,0)[l]{$+$}}
\put(48.00,2.00){\circle{3.00}}
\put(51.00,2.00){\circle*{100.00}}
\put(54.00,2.00){\circle{3.00}}
\put(51.00,5.00){\circle{3.00}}
\put(61.00,2.00){\makebox(0,0)[l]{$+~\dots$}}
\end{picture}
\end{center}
\caption{\it\baselineskip=12pt Resummation of the genus expansion in the
pinched approximation. The solid circles are worldsheet discs (or spheres) and
the thin lines are  strips attached to them with infinitesimal pinching size
$\delta$. Each strip corresponds to an insertion of a bilocal operator on the
worldsheet.}
\bigskip
\label{pinchsum}\end{figure}
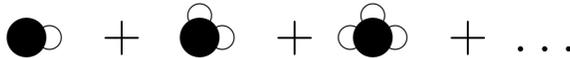

The quantum field theory is described by the fixed-genus Euclidean path
integral
\beq
{\cal Z}_{\rm QFT}[\{g^I\}]=\int D\Phi~\e^{-{\cal L}_{\rm
QFT}[\Phi;\{g^I\}]}~~~~~~,~~~~~~{\cal L}_{\rm QFT}[\Phi;\{g^I\}]={\cal
L}_*[\Phi]+\int_\Sigma d^2z~g^IO_I[\Phi]
\label{QFTdeform}\eeq
where $\Phi$ denotes a collection of fields defined on a Riemann surface
$\Sigma$. The action ${\cal L}_*[\Phi]$ defines a conformal field theory and
$O_I[\Phi]$ are a set of local deformation vertex operators. The sum over all
topologies of the two-dimensional quantum field theory \eqn{QFTdeform} can be
evaluated exactly in the dilute wormhole gas approximation (fig.
\ref{pinchsum}) and it induces statistical fluctuations $\Delta g^I$ of the
$\sigma$-model couplings,
\beq
\sum_{\rm genera}{\cal Z}_{\rm QFT}[\{g^I\}]\simeq\int D\alpha_I~{\cal
P}[\{\alpha_I\}]\int D\Phi~\e^{-{\cal L}_{\rm QFT}[\Phi;\{g^I+\Delta
g^I(\alpha)\}]}
\label{generasumgen}\eeq
where the fields $\alpha_I$ are wormhole parameters. For a dilute gas, the
wormhole probability distribution function is given by
\beq
{\cal P}[\{\alpha_I\}]={\cal N}\exp-\frac1{2\Gamma^2}\,\alpha_I\,{\cal
G}^{IJ}\,\alpha_J
\label{probfngen}\eeq
where $\Gamma$ is the width of the distribution and ${\cal G}^{IJ}$ is an
appropriate metric on the moduli space of coupling constants $\{g^I\}$. This
promotes the couplings $g^I$ to quantum operators $\widehat{g}^I=g^I+\Delta
g^I$ on target space.

Let us now specialize to the case of a system of $N$ non-relativistic heavy
D-particles. In this case, the couplings $\{g^I\}$ are given by the set
$\{Y_{ab}^i(X^0),P_j^{cd}(X^0)\}$, where $Y$ are the collective coordinates of
the D-particles and $P$ are their collective momenta. The indices $i,j$ label
the directions in spacetime while $a,b,c,d=1,\dots,N$ label the $U(N)$ isospin
gauge symmetry present in any multi-brane system. The field $X^0$ is the
worldsheet temporal embedding coordinate and the momenta are given by
$P_i^{ab}=M_sU_i^{ab}$, where\beq
M_s=\frac1{g_s\ell_s}
\label{BPSmass}\eeq
is the mass of a D-particle, with $g_s$ the string coupling constant. According
to the above general prescription, the multi-brane dynamics will induce
position-momentum (or phase space), space-space, and space--time
indeterminancies as follows. From the associated wormhole distribution
\eqn{probfngen} there follows a set of position uncertainties $\Delta Y^i_{ab}$
corresponding to the effective widths $\Gamma^2/G_{IJ}$, where $\Gamma$ is
proportional to the string coupling $g_s$ and $G_{IJ}=\langle V_IV_J\rangle$ is
the Zamolodchikov metric \cite{zam} with $\{V_I\}$ the set of local vertex
operators associated with the D-particle couplings $\{g^I\}$. The momentum
uncertainties $\Delta P^{ab}_j$ arise upon canonical quantization in the moduli
space $\mod$ of $\sigma$-model couplings $\{g^I\}$ which leads to the quantum
commutator
\beq
\left[\!\left[\widehat{Y}^i_{ab},\widehat{P}_j^{cd}\right]\!\right]=i\,
\hbar_{\mods}\,\delta_j^i\,\delta_a^c\,\delta_b^d~~~~~~,~~~~~~
\widehat{P}^i_{ab}=-i\,\hbar_{\mods}\,\frac{\delta}{\delta Y^i_{ab}}
\label{cancomm}\eeq
where the effective Planck constant $\hbar_{\mods}$ of $\mod$ is of order
$g_s$. Furthermore, the target space time $T$ in the ``physical'' frame depends
on the positions and momenta of the D-particles and therefore becomes a target
space operator $\widehat{T}$ upon summation over all worldsheet topologies.
{}From this we will derive a space--time uncertainty relation of the form
\cite{liyoneya} $\Delta Y^i_{ab}\Delta T~\geqsim~{\cal O}(g_s\ell_s^2)$ which
will imply, in particular, that extremely heavy D-particles can probe very
small distances.

\newsection{String Solitons and Non-abelian D-particle Dynamics}

\subsubsection*{Short-distance Spacetime Structure}

The discovery \cite{dbranes,polchinski} of new solitonic structures in
superstring theory have dramatically changed our understanding of target space
structure. These new non-perturbative objects are known as Dirichlet-branes and
they can be seen to arise from the implementation of $T$-duality as a canonical
transformation in the path integral \cite{dornotto,lozano} for the usual open
superstring with free endpoints. The latter object is described by imposing
Neumann boundary conditions on the embedding fields $X^M$, $M=0,1,\dots,9$, of
the string worldsheet $\Sigma$, which we assume has the topology of a disc. At
the circular boundary of $\Sigma$ the fields are constant along the normal
directions, $\partial_{\rm n}X^M=0$, and are allowed to vary as arbitrary
functions $X^M(s)$ on $\partial\Sigma$. Here $\partial_{\rm n}$ is the normal
derivative to the boundary $\partial\Sigma$ in $\Sigma$. A $T$-duality
transformation $X\to\tilde X$, defined on the worldsheet by $\partial_\alpha
X^M=\epsilon^{\alpha\beta}\partial_\beta\tilde X^M$, maps the Neumann boundary
conditions into the
Dirichlet boundary conditions $\partial_{\rm t}\tilde X^M=0$, or equivalently
$\tilde X^M|_{\partial\Sigma}=Y^M$, where $\partial_{\rm t}$ is the derivative
tangent to $\partial\Sigma$. Now the fields are fixed at the specified values
$Y^M$ on $\partial\Sigma$ but can vary in the normal directions. If the
$T$-duality mapping is applied to $9-p$ spatial directions, then the Dirichlet
conditions define a hypersurface in 10-dimensional spacetime. The hypersurface
is embedded into target space from an effective $p+1$ dimensional worldvolume
in which the embedding fields $X^M$ are allowed to vary freely. These objects
are known as D$p$-branes. They are solitons of the open superstring theory and
supersymmetry guarantees their stability. They have a BPS mass given by
\eqn{BPSmass} and are characterized as being topological defects which are
fixed in the $9-p$ spacetime directions. Open string excitations can attach
themselves to the D-brane domain walls.

In this paper we shall specialize to the case of D0-branes, or D-particles. In
this case the set of string embedding fields can be written as $X^M=(X^0,X^i)$,
where $X^0$ is the worldline time coordinate, which satisfies Neumann boundary
conditions, and $X^i$, $i=1,\dots,9$, are the coordinates of the D-particle
which obey Dirichlet boundary conditions. The dynamics of these excitations can
be described by deforming the usual free string $\sigma$-model action $S_0[X]$
by a worldsheet boundary vertex operator \cite{dbranes}
\beq
{\cal S}_{\rm
C}=S_0[X]-\frac1{\ell_s^2}\oint_{\partial\Sigma}ds~Y_i(X^0)\,\partial_{\rm
n}X^i(s)~~~~~~;~~~~~~S_0[X]=\frac1{2\ell_s^2}\int_\Sigma d^2z~\partial
X^M\,\bar\partial X^N\,\eta_{MN}
\label{sigmamodel}\eeq
where $\eta_{MN}$ is a (critical) flat Minkowski spacetime metric. The
non-relativistic motion of heavy D-particles can be described by the
Galilean-boosted configurations $Y^i(X^0)=Y^i+U^iX^0$ where $U^i$ is the
non-relativistic velocity of the branes.

\begin{figure}[htb]
\epsfxsize=1.5in
\bigskip
\centerline{\epsffile{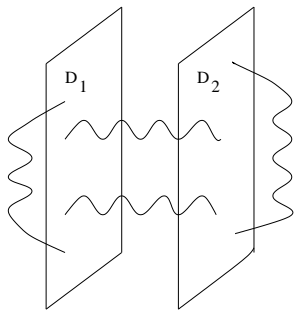}}
\caption{\it\baselineskip=12pt Emergence of the enhanced $U(N)$ gauge symmetry
for bound states of $N$ parallel D-branes (planes). An oriented
fundamental string (wavy lines) can start and end either at the same or
different D-brane, giving $N^2$ massless vector states in the limit of
coinciding branes. These states form a representation of $U(N)$.}
\bigskip
\label{dbrane}\end{figure}

The interesting situation arises when one considers a configuration of $N$
D-branes (fig. \ref{dbrane}). The multiple D-brane assembly leads to a
non-commutative structure at very short distances in the spacetime. The
situation is actually quite simple. Consider a system of $N$ parallel D-branes.
When the separation between branes is of a sub-Planckian distance scale, the
solitons can interact with each other via the exchange of open strings. It is
this property of D-brane dynamics that makes them good probes of the
short-distance structure of spacetime and it implies that spacetime at very
small distance scales is described by some sort of non-commutative geometry. A
tractable limit of this situation is the case of overlapping branes or
infinitesimal separations. More complicated situations can also arise, for
instance when the branes interact via the exchange of other D-branes, in
addition to their string interactions. One would then need to employ a
formalism for intersecting D-brane configurations. In the following we will
concentrate on the simpler situation of only open string excitations between
the D-branes.

\subsubsection*{Low-energy Effective Field Theories}

To understand why a non-commutative structure is implied at very short distance
scales, one needs to examine the effective field theory description for $N$
parallel D-branes. In the multi-D-particle case, the assembly is described by a
set of $N\times N$ Hermitian matrices $Y^i_{ab}$, $i=1,\dots,9$, in the adjoint
representation of the unitary group $U(N)$ which are obtained as the remnant
fields from the dimensional reduction of 10-dimensional maximally
supersymmetric $U(N)$ Yang-Mills theory to the worldlines of the D-particles
\cite{bound}. The reduced Yang-Mills potential
\beq
V_0[Y]=\frac1{\ell_s^4}\frac1{4g_s}\sum_{i,j=1}^9\tr\left[Y^i,Y^j\right]^2
\label{YMpotential}\eeq
then governs the dynamics of the branes, where tr denotes the trace in the
fundamental representation of $U(N)$. There are two limiting regimes of this
theory. In the weak-coupling limit $g_s\to0$, the branes are very far apart and
do not interact with each other. The potential \eqn{YMpotential} is minimized
by those matrix-valued configurations which satisfy $[Y^i,Y^j]=0$, $\forall
i,j$. These solutions correspond to states of maximal supersymmetry. In
this case the generic $U(N)$ gauge symmetry of the Yang-Mills theory is broken
down to $U(1)^N$, i.e. there is one $U(1)$ gauge field on each D-brane.
Moreover, the matrix fields can be simultaneously diagonalized by a gauge
transformation to give $Y^i={\rm diag}\{y_{(a)}^i\}_{a=1,\dots,N}$, where the
eigenvalues $y_{(a)}^i$ represent the coordinates of each D-particle. Now
consider the opposite case where the branes are almost on top of each other.
Since the energy of a fundamental string which stretches between two different
branes is $\frac1{2\ell_s}|\vec y_{(a)}-\vec y_{(b)}|$, it follows that in this
case more massless vector states appear in the spectrum of the $U(N)$ gauge
theory. Now the configurations of the Yang-Mills potential \eqn{YMpotential}
satisfy $[Y^i,Y^j]\neq0$ for $i\neq j$ and correspond to states of broken
supersymmetry which
must be incorporated into the quantum gauge theory. Thus the limit of
coinciding branes restores the full $U(N)$ gauge symmetry and leads to a Lie
algebraic non-commutativity in the spacetime coordinates $Y^i_{ab}$. The
components with $a\neq b$ are to be interpreted as the coordinates of the short
open string degrees of freedom stretched between the branes, so that the
D-brane coordinates are viewed as adjoint Higgs fields in this picture. These
ideas are depicted schematically in fig. \ref{dbrane}.

In the following we shall be interested in establishing the manner in which
this Lie algebraic non-commutativity implies a genuine quantum spacetime
non-commutativity. For this, we consider an alternative description to the
Yang-Mills matrix quantum mechanics which is given by a non-local deformation
of a free worldsheet $\sigma$-model. In the simplest case of non-relativistic
motion, it follows from the BPS mass formula \eqn{BPSmass} that the limit of
heavy D-particles is equivalent to taking $g_s\ll1$. This is precisely the
regime in which worldsheet perturbation theory can be trusted. The partition
function for the uniform motion of the multi-D0-brane system in the $T$-dual
Neumann picture is defined as the expectation value of a path-ordered Wilson
loop operator along the worldsheet boundary in the free $\sigma$-model,
\beq
Z_N[Y]=\int DX~\e^{-S_0[X]}~\tr\,{\rm
P}\,\exp\left(ig_s\oint_{\partial\Sigma}A_M(X^0(s))~dX^M(s)\right)
\label{NCaction}\eeq
where the field $A$ is an $N\times N$ Hermitian matrix in the adjoint
representation of $U(N)$ with $A^M=(A^0,-\frac1{\ell_s^2}Y^i)$ the components
of the $U(N)$ gauge potential dimensionally reduced to the worldline of the
D-particles.

\begin{figure}[htb]
\epsfxsize=2.5in
\bigskip
\centerline{\epsffile{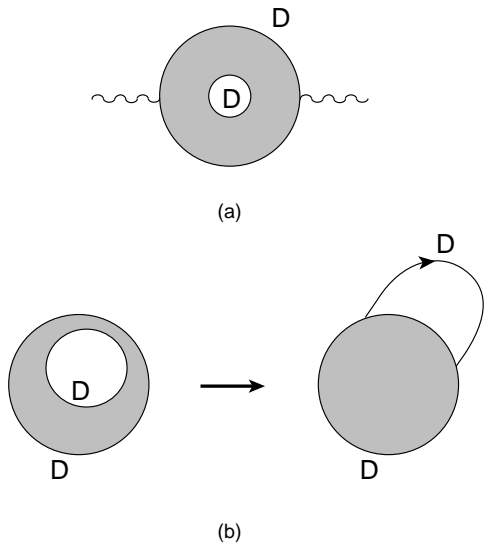}}
\caption{\baselineskip=12pt (a) {\it World-sheet annulus diagram for the
leading quantum correction to the propagation of a string state $V$ (wavy
lines) in a D-brane background, and} (b) {\it the pinched annulus configuration
which is the dominant divergent contribution to the quantum recoil.}}
\bigskip
\label{annulus}\end{figure}

However, the model \eqn{NCaction} on its own is a bit too naive and needs to be
supplemented with some auxilliary prescriptions. There are problems with
summing over worldsheet genera in the Dirichlet picture, which could be related
to the breaking of the $T$-duality symmetry from an anomaly in the non-abelian
case \cite{dornlast,brecher}. Modular logarithmic divergences appear in matter
field amplitudes when the string propagator $L_0$ is computed with Dirichlet
boundary conditions. Consider a string matter field polarization tensor $V$ in
the D-brane background at the level of an annular topology (fig.
\ref{annulus}). Corresponding to a string state of conformal weight $\Delta$,
the string propagator contains a modular parameter integration of the form
$\int\frac{dq}q~q^\Delta$. In the pinched annulus limit, with infinitesimal
pinching size $\delta$, the states of zero conformal dimension $\Delta=0$
therefore yield logarithmic divergences of the form $\log\delta$.

These modular divergences are cancelled by adding logarithmic recoil operators
\cite{kmw} to the matrix $\sigma$-model action \eqn{NCaction}. If one is to use
low-energy probes to observe short-distance spacetime structure, such as a
generalized Heisenberg microscope, then one needs to consider the scattering of
string matter off the D-particles. The worldsheet perspective of this physical
situation is represented by fig. 3. The target space picture is the following.
At time $t<0$ a closed string state propagates towards the spacetime string
defect which is fixed in space. At $t=0$ it interacts with the defect by
splitting and attaching itself to the D-particle. Then at time $t>0$, the
instantaneous interaction causes a transfer of energy from the string state to
the defect such that the D-particle recoils with some non-zero velocity. For
the Galilean-boosted multi-D-particle system, the recoil is described by taking
the deformation of the $\sigma$-model action in \eqn{NCaction} to be of the
form \cite{ms}
\beq
Y^i_{ab}(X^0)=\lim_{\epsilon\to0^+}\left(\ell_sY^i_{ab}\,C_\epsilon(X^0)+
U^i_{ab}\,D_\epsilon(X^0)\right)
\label{recoildef}\eeq
where
\beq
C_\epsilon(X^0)=\epsilon\,\theta_\epsilon(X^0)~~~~~~,~~~~~~
D_\epsilon(X^0)=X^0\,\theta_\epsilon(X^0)
\label{recoilops}\eeq
and
\beq
\theta_\epsilon(X^0)=\frac1{2\pi
i}\int_{-\infty}^{+\infty}\frac{dq}{q-i\epsilon}~\e^{iqX^0}
\label{stepfnreg}\eeq
is the regulated step function whose $\epsilon\to0^+$ limit is the usual step
function. The operators \eqn{recoilops} have non-vanishing matrix elements
between different string states and therefore describe the appropriate change
of state of the D-brane background. They can be thought of as describing the
recoil of the assembly of D-particles in an impulse approximation, in which it
starts moving as a whole only at time $X^0=0$. The collection of constant
matrices $\{Y^i_{ab},U^j_{cd}\}$ now form the set of coupling constants for the
worldsheet $\sigma$-model \eqn{NCaction}.

\subsubsection*{Galilean Invariance and Worldsheet Logarithmic Conformal
Algebra}

The recoil operators \eqn{recoilops} possess a very important property. They
lead to a deformation of the free $\sigma$-model action in \eqn{NCaction} which
is not conformally-invariant, but rather defines a logarithmic conformal field
theory \cite{gurarie}. Logarithmic conformal field theories lie on the border
between conformal field theories and generic two-dimensional renormalizable
field theories. They contain logarithmic scaling violations in their
correlation
functions on the worldsheet. In the present case, this can be seen by computing
the pair correlators of the fields \eqn{recoilops} which give \cite{kmw}
\beq
\Bigl\langle
C_\epsilon(z)\,C_\epsilon(0)\Bigr\rangle=0~~~~~~,~~~~~~\Bigl\langle
C_\epsilon(z)\,D_\epsilon(0)\Bigr\rangle=\frac
b{z^{\Delta_\epsilon}}~~~~~~,~~~~~~\Bigl\langle
D_\epsilon(z)\,D_\epsilon(0)\Bigr\rangle=\frac{b\,\ell_s^2}
{z^{\Delta_\epsilon}}\,\log z
\label{2ptconfalg}\eeq
where
\beq
\Delta_\epsilon=-\frac{|\epsilon|^2\,\ell_s^2}2
\label{CDconfdim}\eeq
is the conformal dimension of the recoil operators. The constant $b$ is fixed
by the leading logarithmic divergence of the conformal blocks of the theory.
Note that \eqn{CDconfdim} vanishes as $\epsilon\to0$, so that the logarithmic
worldsheet divergences in \eqn{2ptconfalg} cancel the modular annulus
divergences discussed above. An essential ingredient for this cancellation is
the identification \cite{kmw}
\beq
\epsilon^{-2}=-2\ell_s^2\log\Lambda
\label{scalerel}\eeq
which relates the target space regularization parameter $\epsilon$ to the
worldsheet ultraviolet cutoff scale $\Lambda$ (the minus sign in \eqn{scalerel}
is due to the Minkowski signature of $X^0$).

Logarithmic conformal field theories are characterized by the fact that their
Virasoro generator $L_0$ is not diagonalizable, but rather admits a Jordan cell
structure. Here the operators \eqn{recoilops} form the basis of a $2\times2$
Jordan block and they appear in the spectrum of the two-dimensional quantum
field theory as a consequence of the zero modes that arise from the breaking of
the target space translational symmetry by the topological defects. The mixing
between $C$ and $D$ under a conformal transformation of the worldsheet can be
seen explicitly by considering a finite-size scale transformation
\beq
\Lambda\to\Lambda'=\Lambda\,\e^{-t/\ell_s}
\label{Lambdatransf}\eeq
Using \eqn{scalerel} it follows that the operators \eqn{recoilops} are changed
according to $D_\epsilon\to D_\epsilon+t\ell_sC_\epsilon$ and $C_\epsilon\to
C_\epsilon$. Thus in order to maintain scale-invariance of the theory
\eqn{NCaction} the coupling constants must transform under \eqn{Lambdatransf}
as \cite{lm,kmw} $Y^i\to Y^i+U^it$ and $U^i\to U^i$, which are just the
Galilean transformation laws for the positions $Y^i$ and velocities $U^i$. Thus
a finite-size scale transformation of the worldsheet is equivalent to a
Galilean transformation of the moduli space of $\sigma$-model couplings, with
the parameter $\epsilon^{-2}$ identified with time $t$. The corresponding
$\beta$-functions for the worldsheet renormalization group flow are
\beq
\beta_{Y_i}\equiv\frac{dY_i}{dt}=\Delta_\epsilon\,Y_i+\ell_s\,U_i
{}~~~~~~,~~~~~~\beta_{U_i}\equiv\frac{dU_i}{dt}=\Delta_\epsilon\,U_i
\label{betafns}\eeq

\newsection{Quantization of Moduli Space}

We shall now begin describing the basic steps towards the quantization of the
$\sigma$-model couplings representing the collective degrees of freedom of the
assembly of D-particles.

\subsubsection*{Liouville-dressed Renormalization Group Flows}

We shall first need to identify the time variable of our system. Note that for
finite $\epsilon$, \eqn{CDconfdim} shows that the operators \eqn{recoilops}
lead to a relevant deformation of the free $\sigma$-model. The deformation
becomes marginal in the limit $\epsilon\to0$. When the field theory lies away
from criticality we must dress the model by Liouville theory \cite{ddk} in
order to restore conformal invariance at the quantum level. The worldsheet zero
mode of the Liouville field can then be identified with the local worldsheet
regularization scale. Thus the Liouville field is interpreted as the target
space time \cite{time}, and from the discussion of the previous section, it
coincides with the temporal embedding field $X^0$. This means that the
incorporation of the regulated operators (\ref{recoilops}, \ref{stepfnreg}) can
be thought of as the appropriate dressing of the bare coupling constants
$\{Y^i_{ab},U^j_{cd}\}$ of the $\sigma$-model \cite{diffusion,ms}.

In general, the Liouville dynamics ensures the possibility of canonical
quantization in the moduli space of $\sigma$-model couplings through a set of
properties known as Helmholtz conditions \cite{ms,emn}. The Liouville field
$\phi$ is defined by identifying the conformal equivalence class of the metric
$\gamma_{\alpha\beta}$ of $\Sigma$,
\beq
\gamma_{\alpha\beta}=\e^{(2/\ell_sQ)\phi}\,\widehat{\gamma}_{\alpha\beta}
\label{phidef}\eeq
where $\widehat{\gamma}_{\alpha\beta}$ is a fixed fiducial worldsheet metric
and $Q$ is related to the central charge of the corresponding two-dimensional
quantum gravity. Then the Liouville dressing of the deformation of a free
$\sigma$-model action $S_0$, which is characterized by a set of vertex
operators $\{V_I\}$ with corresponding coupling constants $\{g^I\}$, is
described by the action
\beq
S_\sigma^{\rm(L)}=S_0+\int_\Sigma
d^2z~g^I(\phi)\,V_I+\frac1{2\ell_s^2}\int_\Sigma
d^2z~\partial\phi\,\bar\partial\phi-\frac Q{2\ell_s^2}\int_\Sigma
d^2z~\phi\,R^{(2)}-\frac Q{2\ell_s^2}\oint_{\partial\Sigma}d\widehat{s}~\phi\,K
\label{liouvilleaction}\eeq
where $R^{(2)}$ is the scalar curvature of $\Sigma$ and $K$ is the extrinsic
curvature at the boundary $\partial\Sigma$ ($K=2$ for a disc). From
\eqn{phidef} it follows that the Liouville zero mode is
$\phi_0=-\ell_sQ\log\Lambda$. Microscopically, quantum fluctuations in the
dressed variables $g^I(\phi)$ are induced by summing over all worldsheet
topologies, in analogy with the Coleman approach \cite{wormholes} to wormhole
calculus and the quantization of coupling constants in quantum gravity
\cite{emn}.

\subsubsection*{D-particle Dynamics on Moduli Space}

We are now ready to present the formalism for multi-D-particle dynamics. In the
following we shall, for simplicity, concentrate only on the case of the
constituent motion of the particles, i.e. we will subtract out their center of
mass degree of freedom $Y_i^{\rm cm}=\frac1N\,\tr\,Y_i$. This means that the
effective
gauge symmetry of the D-particle system is now $SU(N)$. To write the partition
function \eqn{NCaction} in the form of a local deformation of the free
$\sigma$-model action $S_0[X]$, we need to disentangle the path-ordering in the
Wilson loop operator. This is done by considering a set of corresponding
abelianized vertex operators which are defined using an auxilliary field
formalism for the $SU(N)$-invariant theory \cite{ms,dornotto,dornlast,dorn}. We
introduce a set of complex auxilliary fields $\bar\xi_a(s),\xi_b(s)$ which live
on the boundary $\partial\Sigma$ of the worldsheet and whose propagator is
$\langle\bar\xi_a(s)\xi_b(s')\rangle=\delta_{ab}\theta(s'-s)$. The partition
function \eqn{NCaction} can then be written as
\bea
Z_N[Y]&=&\int DX~\e^{-S_0[X]}~\frac1N\int
D\bar\xi~D\xi~\bar\xi_c(0)\,\xi_c(1)\nn\\&
&\times\exp\left(-\int_0^1ds~\left[\bar\xi_a(s)\,\frac
d{ds}\xi_a(s)-\frac{ig_s}{\ell_s^2}\,\bar\xi_a(s)\,Y_i^{ab}(X^0(s))\,
\frac{dX^i(s)}{ds}\,\xi_b(s)\right]\right)\nn\\& &
\label{NCauxrep}\eea
in the static gauge $A_0=0$. If we leave the integration over auxilliary fields
in \eqn{NCauxrep} until the very end, then the partition function is expressed
as a functional integral involving the local action
\beq
{\cal S}=S_0[X]+\oint_{\partial\Sigma}ds~Y_i^{ab}(X^0(s))\,V_{ab}^i(X;s)
\label{NCabelianaction}\eeq
where the deformation is described by the set of vertex operators
\beq
V_{ab}^i(X;s)=-\frac{ig_s}{\ell_s^2}\left(\frac{dX^i(s)}{ds}\right)
\bar\xi_a(s)\xi_b(s)
\label{abvertexops}\eeq
The action \eqn{NCabelianaction} is the appropriate non-abelian version of (the
$T$-dual of) the worldsheet action \eqn{sigmamodel} describing the dynamics of
a single D-brane, and as such it represents the abelianization of the
non-abelian D-particle dynamics.

The Zamolodchikov metric $G_{ab;cd}^{ij}=2N\Lambda^2\langle
V_{ab}^i(X;0)V_{cd}^j(X;0)\rangle$ on the moduli space $\mod$ controls the
dynamics of the D-particles, where the vacuum expectation value is taken with
respect to the partition function \eqn{NCauxrep}. This two-point function can
be evaluated to leading orders in $\sigma$-model perturbation theory using the
logarithmic conformal algebra \eqn{2ptconfalg} and the propagator of the
auxilliary fields to give \cite{ms}
\bea
G_{ab;cd}^{ij}&=&\frac{4\bar g_s^2}{\ell_s^2}\left[\eta^{ij}\,I_N\otimes
I_N+\frac{\bar g_s^2}{36}\left\{I_N\otimes\left(\bar U^i\bar U^j+\bar U^j\bar
U^i\right)\right.\right.\nn\\& &\biggl.\left.+\,\bar U^i\otimes\bar U^j+\bar
U^j\otimes\bar U^i+\left(\bar U^i\bar U^j+\bar U^j\bar U^i\right)\otimes
I_N\right\}\biggr]_{db;ca}+{\cal O}\left(\bar g_s^6\right)
\label{zammetricexpl}\eea
where $I_N$ is the identity operator of $SU(N)$ and we have introduced the
renormalized coupling constants $\bar g_s=g_s/\ell_s\epsilon$ and $\bar
U^i=U^i/\ell_s\epsilon$. From the renormalization group equations \eqn{betafns}
it follows that the renormalized velocity operator in target space is truly
marginal, $\frac{d\bar U^i}{dt}=0$, which ensures uniform motion of the
D-branes. It can also be shown that the renormalized string coupling $\bar g_s$
is time-independent \cite{ms}. If we further define the position
renormalization $\bar Y^i=Y^i/\ell_s\epsilon$, then the $\beta$-function
equations \eqn{betafns} coincide with the equations of motion of the
D-particles, i.e. $\frac{d\bar Y^i}{dt}=\bar U^i$. Note that the Zamolodchikov
metric \eqn{zammetricexpl} is a complicated function of the D-brane dynamical
parameters. It therefore represents the appropriate effective target space
geometry of the D-particles and, as we will see, it naturally encodes the
short-distance properties of the D-particle spacetime. The canonical momentum
$P_{ab}^i$ for the D-particle dynamics on moduli space can also be determined
perturbatively in the $\sigma$-model \eqn{NCauxrep} by noting that the
Schr\"odinger representation of the Heisenberg algebra \eqn{cancomm} implies
that it is the one-point function of the deformation vertex operators,
$P_{ab}^i=\left\langle V_{ab}^i(X;0)\right\rangle$. A long and tedious
calculation using the three-point correlation functions of the logarithmic pair
\cite{ms} gives
\beq
P_{ab}^i=\frac{8\bar g_s^2}{\ell_s}\left[\bar U^i+\frac{\bar g_s^2}6\left(\bar
U_j^2\bar U^i+\bar U_j\bar U^i\bar U^j+\bar U^i\bar
U_j^2\right)\right]_{ba}+{\cal O}\left(\bar
g_s^6\right)=\ell_s\,G_{ab;cd}^{ij}\,\dot{\bar Y}_{\!j}^{cd}
\label{canmompert}\eeq
which, as expected for the uniform D-particle motion here, coincides with the
contravariantized velocity on $\mod$.

We can now write down an associated effective target space Lagrangian which is
defined in terms of the standard non-linear $\sigma$-model on $\mod$,
\beq
{\cal L}_\mods=-\mbox{$\frac{\ell_s}2$}\,\dot{\bar
Y}_{\!i}^{ab}\,G_{ab;cd}^{ij}\,\dot{\bar Y}_{\!j}^{cd}+\dots
\label{modaction}\eeq
where the dots denote potential terms involving the Zamolodchikov $C$-function
\cite{zam} plus additional terms which depend on the choice of renormalization
scheme. The Lagrangian \eqn{modaction} is readily seen to coincide with the
expansion to ${\cal O}(\bar g_s^4)$ of the symmetrized form of the non-abelian
Born-Infeld action for the D-brane dynamics \cite{brecher,tseytlin},
\beq
{\cal L}_{\rm NBI}=\frac1{\ell_s\bar g_s}\,\tr~{\rm
Sym}\,\sqrt{\det_{M,N}\left[\eta_{MN}\,I_N+\ell_s^2\bar g_s^2\,F_{MN}\right]}
\label{NBIaction}\eeq
where ${\rm Sym}(M_1,\dots,M_n)=\frac1{n!}\sum_{\pi\in S_n}M_{\pi_1}\cdots
M_{\pi_n}$ is the symmetrized matrix product and the components of the
dimensionally reduced field strength tensor are given by
$F_{0i}=\frac1{\ell_s^2}\dot{\bar Y}_{\!i}$ and $F_{ij}=\frac{\bar
g_s}{\ell_s^4}[\bar Y_i,\bar Y_j]$. In the abelian reduction to the case of a
single D-particle, the Lagrangian \eqn{NBIaction} reduces to the usual one
describing the free relativistic motion of a massive particle. The leading
order $F^2$ term in the expansion of \eqn{NBIaction} is just the usual
Yang-Mills Lagrangian. The formalism described here thereby represents a highly
non-trivial application of the theory of logarithmic operators.

Finally, we come to the definition of the target space time. The flat
worldsheet Zamolodchikov $C$-theorem \cite{zam} can be expressed as
\beq
\frac{\partial{\cal C}}{\partial t}=-\,\e^{-{\cal
C}t/\ell_s}\,\beta_i^{ab}\,G_{ab;cd}^{ij}\,\beta_j^{cd}
\label{zamthm}\eeq
where the running central charge ${\cal C}[Y]\sim Q^2$ is the Zamolodchikov
$C$-function and the exponential factor in \eqn{zamthm} comes from the
extrinsic curvature term in the Liouville-dressed action \eqn{liouvilleaction}.
The non-linear differential equation \eqn{zamthm} can be solved for small
velocities (extreme non-relativistic motion) to give the physical target space
time coordinate \cite{ms}
\beq
T\equiv\sqrt{\cal C}\,t\simeq\frac{2\bar g_st}{\sqrt{\ell_s}}\,\sqrt{{\cal
U}(\bar U)\int_0^td\tau~\e^{2(\tau^2-t^2)\bar g_s^2{\cal U}(\bar U)/\ell_s^2}}
\label{phystime}\eeq
where we have introduced the velocity-dependent invariant function
\beq
{\cal U}(\bar U)=\tr\,\bar U_i^2+\frac{\bar g_s^2}{36}\,\tr\left(2\bar
U_i^2\bar U_j^2+\bar U_i\bar U_j\bar U^i\bar U^j\right)+{\cal O}\left(\bar
g_s^4\right)
\label{calUdef}\eeq
The definition \eqn{phystime} comes from the normalization of the Liouville
field kinetic term in \eqn{liouvilleaction} appropriate to a Robertson-Walker
spacetime geometry \cite{time}. Its expression in \eqn{phystime} holds near any
fixed point in moduli space and is valid in the usual regime of applicability
of worldsheet $\sigma$-model perturbation theory.

\subsubsection*{The Genus Expansion}

We shall now describe the process of coupling constant quantization via the sum
over worldsheet topologies for the model \eqn{NCaction}. The key point in the
non-abelian case is that the sum over genera and the auxilliary field
representation of the Wilson loop operator in \eqn{NCauxrep} commute, allowing
one to write
\bea
\sum_{\rm genera}Z_N[Y]&=&\sum_{h=0}^\infty\int DX~\e^{-S_0^{(h)}[X]}~\tr\,{\rm
P}\,\exp\left(\frac{ig_s}{\ell_s^2}\oint_{\partial\Sigma_h}A_M(X^0)~dX^M
\right)\nn\\&=&\frac1N\int D\bar\xi~D\xi~\bar\xi_c(0)\,\xi_c(1)\nn\\& &
\times\sum_{h=0}^\infty\int DX~\exp\left(-S_0^{(h)}[X]-
\oint_{\partial\Sigma_h}ds~Y_i^{ab}(X^0)\,V_{ab}^i(X;s)\right)
\label{genusexp}\eea
where $S_0^{(h)}[X]$ is the free $\sigma$-model action defined on a genus $h$
Riemann surface $\Sigma_h$ with $h+1$ boundaries (so that $\partial\Sigma_h$ is
a disjoint union of $h+1$ circles). Therefore, if we again leave the auxilliary
field
integrations until the very end, then we can exploit the abelianization of the
non-abelian dynamics in \eqn{genusexp} to study the topological expansion.

The latter quantity can be described precisely in the pinched approximation
(fig. \ref{pinchsum}). There are two sorts of modular divergences which
dominate this truncation of the string genus expansion. The leading ones are of
the form $(\log\delta)^2$ and arise from the logarithmic nature of the
deformation. It can be shown \cite{ms} that these divergences are cancelled by
requiring that the velocities of the D-particles in the scattering of string
matter off them change according to $\Delta \bar U_i^{ab}=-\frac1{M_s}(k_{\rm
i}+k_{\rm f})_i\delta^{ab}$, where $M_s=1/\bar g_s\ell_s$ is the BPS mass of
the string solitons and $k_{\rm i,f}$ denote the initial and final momenta in
the scattering process. Thus the leading divergences of the genus expansion are
cancelled by imposing momentum conservation in scattering processes involving
string matter. Note that this result only controls the dynamics of the
constituent D-branes themselves and not the open string excitations connecting
them. It therefore tells us nothing about short-distance (non-commutative)
spacetime structure.

This latter property of the moduli space comes from examining the sub-leading
divergences, which are of the form $\log\delta$ and are associated with the
vanishing conformal dimension of the logarithmic operators. The regularization
of these singularities induces quantum fluctuations of the D-particle
collective coordinates and leads to short-distance uncertainties. In the
pinched approximation represented by fig. \ref{pinchsum}, the effect of the
dilute gas of wormholes is to exponentiate the bilocal operator inserted on the
boundary of the disc $\Sigma=\Sigma_0$. This leads to a change of the action
\eqn{NCabelianaction} given by
\beq
\Delta{\cal
S}\simeq\frac{g_s^2}2\,\log\delta\,\oint\!\!\oint_{\partial\Sigma}ds~ds'~
V_{ab}^i(X;s)\,G^{ab;cd}_{ij}\,V_{cd}^j(X;s')
\label{actionchange}\eeq
This bilocal interaction term can be written as a local worldsheet effective
action by using standard tricks of wormhole calculus \cite{wormholes} and
introducing wormhole parameters $\rho_i^{ab}$ on the moduli space $\mod$ which
linearize \eqn{actionchange} via a functional Gaussian integral transformation.
The net result of the summation over genera in the pinched approximation is
therefore
\bea
\sum_{\rm genera}Z_N[Y]&\simeq&\int_\mods
D\rho~\exp\left(-\frac1{2\Gamma^2}\,\rho_i^{ab}\,G_{ab;cd}^{ij}\,
\rho_j^{cd}\right)\nn\\& &\times\int DX~\e^{-S_0[X]}~\tr\,{\rm P}\,
\exp\left(\frac{ig_s}{\ell_s^2}\oint_{\partial\Sigma}\left(Y_i(X^0)+
\rho_i(X^0)\right)~dX^i\right)
\label{genussumfinal}\eea
where the width of the Gaussian wormhole distribution function in
\eqn{genussumfinal} is given by $\Gamma=g_s\,\sqrt{\log\delta}$. Eq.
\eqn{genussumfinal} shows that the genus expansion induces statistical
fluctuations $\Delta Y_i^{ab}=g_s\sqrt{\log\delta}\,\rho_i^{ab}$ of the
coordinates $Y_i^{ab}$ of the assembly of D-particles. As discussed in section
1, it is this property that will allow us to probe short-distance spacetime
structure in terms of the geometry and the dynamics on moduli space.

\subsubsection*{Diagonalization of Moduli Space}

To be able to write down a set of position uncertainties for each direction of
target space and of the $SU(N)$ group manifold, we shall need to diagonalize
the bilinear form of the wormhole parameter distribution in
\eqn{genussumfinal}. This requires the diagonalization of the inverse of the
Zamolodchikov metric. The diagonalization of the moduli space $\mod$ reveals
the precise manner in which the string interactions between
D-particles induce short-distance non-commutativity. This leads to a very nice
dynamical and geometrical picture of short distance spacetime structure.

For this, we employ a Born-Oppenheimer approximation to the D-particle
interactions to separate the diagonal D-particle coordinates from the
off-diagonal parts of the adjoint Higgs fields representing the short open
string excitations connecting them. This approximation is valid in the limit of
small velocities \cite{liyoneya} which corresponds to a configuration of
well-separated branes. In the free string limit $g_s\ll1$, we may diagonalize
the configuration fields simultaneously in the static gauge by a
time-independent gauge transformation,
\beq
\bar Y^i=\Omega~{\rm
diag}\left(y_{(1)}^i,\dots,y_{(N)}^i\right)~\Omega^{-1}~~~~~~,~~~~~~\Omega\in
SU(N)
\label{diagonalization}\eeq
where the eigenvalues $y_{(a)}^i\in\real$ are the positions of the D-particles
which move at constant velocities $u_{(a)}^i=dy_{(a)}^i/dt$. The unitary
transformation \eqn{diagonalization} diagonalizes the Zamolodchikov metric
\eqn{zammetricexpl} in its $su(N)\otimes su(N)$ indices and we have
\beq
G^{ij}=\frac{4\bar
g_s^2}{\ell_s^2}\,(\Omega\otimes\Omega)\left[\eta^{ij}\,I_N\otimes
I_N+\frac{\bar g_s^2}{36}\,{\cal U}^{ij}+{\cal O}\left(\bar
g_s^4\right)\right](\Omega\otimes\Omega)^{-1}
\label{zamgpdiag}\eeq
where
\beq
{\cal
U}^{ij}_{ab;cd}=\left(2u_{(a)}^iu_{(a)}^j+2u_{(b)}^iu_{(b)}^j+
u_{(a)}^iu_{(b)}^j+u_{(a)}^ju_{(b)}^i\right)\,\delta_{ad}\,\delta_{bc}
\label{Ugen}\eeq

\begin{figure}[htb]
\epsfxsize=3.5in
\bigskip
\centerline{\epsffile{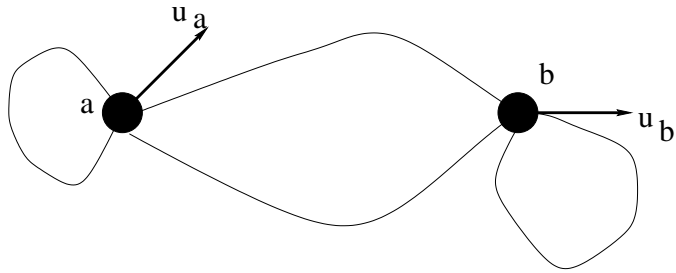}}
\caption{\it\baselineskip=12pt The string frame representation of D-particles
$a$ and $b$ moving at velocities $\vec u_{(a)}$ and $\vec u_{(b)}$ in target
space.}
\bigskip
\label{stringframe}\end{figure}

It now remains to diagonalize the operator \eqn{Ugen} in its $9\times9$
spacetime indices $i,j$. The situation is very simple when $a=b$, as then the
eigenvalues of \eqn{Ugen} are given by
\beq
\lambda_{aa}^1=6\vec
u_{(a)}^{\,2}~~~~~~,~~~~~~\lambda_{aa}^2=\dots=\lambda_{aa}^9=0
\label{eigenaa}\eeq
The orthogonal matrix $O_{aa}$ which diagonalizes the Zamolodchikov metric in
this case is simply the $9\times9$ identity matrix upon rotation to the
coordinate system in which the first direction is spanned by the normalized
velocity vector $u_{(a)}^i/|\vec u_{(a)}|$. We shall refer to this frame as the
``string frame'' as it represents the one-dimensional coordinate system
relative to the single open string excitation that starts and ends on the same
D-particle $a$ (fig. \ref{stringframe}). The situation is far more complicated
for $a\neq b$ because now the string interactions between a given pair of
D-particles $a,b$ also play a role. In this case the eigenvalues are
\bea
& &\lambda_{ab}^{1,2}=\vec u_{(a)}^{\,2}+\vec u_{(b)}^{\,2}+\vec
u_{(a)}\cdot\vec u_{(b)}\nn\\& &\pm\,\sqrt{\left(\vec u_{(a)}^{\,2}+\vec
u_{(b)}^{\,2}+\vec u_{(a)}\cdot\vec u_{(b)}\right)^2+\frac{\left[\left(\vec
u_{(a)}\cdot\vec u_{(b)}\right)^2+\vec u_{(a)}^{\,2}\vec u_{(b)}^{\,2}+2\vec
u_{(a)}\cdot\vec u_{(b)}\left(\vec u_{(a)}^{\,2}+\vec
u_{(b)}^{\,2}\right)\right]^2}{\vec u_{(a)}^{\,2}\vec u_{(b)}^{\,2}-\left(\vec
u_{(a)}\cdot\vec u_{(b)}\right)^2}}\nn\\&
&\lambda_{ab}^3=\dots=\lambda_{ab}^9=0
\label{eigenab}\eea
If we assume that the velocity vectors $\vec u_{(a)}$ and $\vec u_{(b)}$ are
linearly independent, then they span a two-dimensional space which we refer to
as the string plane. The increase in dimension of this frame owes to the
increase in degrees of freedom of the open string which now stretches between
two different branes (fig. \ref{stringframe}). In this coordinate system, the
orthogonal diagonalization matrix is
\bea
O_{ab}&=&\frac1{\left|\vec u_{(a)}+B(u)\vec u_{(b)}\right|^{2/9}}\nn\\&
&\times\pmatrix{\left|\vec u_{(a)}\right|+B(u)\left|\vec
u_{(b)}\right|\cos\theta_{ab}&-B(u)\left|\vec
u_{(b)}\right|\sin\theta_{ab}&0&\dots&0\cr
B(u)\left|\vec u_{(b)}\right|\sin\theta_{ab}&\left|\vec
u_{(a)}\right|+B(u)\left|\vec
u_{(b)}\right|\cos\theta_{ab}&0&\dots&0\cr0&0&1&\dots&0\cr\vdots&\vdots&
\vdots&\ddots&\vdots\cr0&0&0&\dots&1\cr}\nn\\& &
\label{Oab}\eea
where $\theta_{ab}$ is the angle between the velocity vectors and
\beq
B(u)=\frac{\left(\vec u_{(a)}\cdot\vec u_{(b)}\right)^2+\vec u_{(a)}^{\,2}\vec
u_{(b)}^{\,2}+2\vec u_{(a)}\cdot\vec
u_{(b)}\left(\vec u_{(a)}^{\,2}+\vec u_{(b)}^{\,2}\right)-\lambda_{ab}^1\,\vec
u_{(a)}\cdot\vec u_{(b)}}{2\vec u_{(a)}^{\,2}\left(\vec u_{(a)}\cdot\vec
u_{(b)}\right)+2\left(\vec u_{(a)}\cdot\vec u_{(b)}\right)^2+2\vec
u_{(a)}^{\,4}-\lambda_{ab}^1\,\vec u_{(a)}^{\,2}}
\label{Budef}\eeq
It is in this way that the Zamolodchikov metric on $\mod$ naturally captures
the geometry of the string interactions among the D-branes and illustrates the
complexity change between the dynamics of a single D-particle on its own
($a=b$) and the interactions of a multi-brane system ($a\neq b$).

\newsection{Quantum Uncertainty Relations}

The desired position coordinates in which the bilinear form of
\eqn{genussumfinal} is diagonal are given by $\tilde
Y_{ab}^i=O_{ab}^{ji}[\Omega^{*-1}\bar Y_j\Omega]_{ba}\equiv O_{ab}^{ji}{\cal
Y}_j^{ab}(\bar Y)$, where the complex configuration fields ${\cal Y}^i_{ab}$
encode the information about the string interactions between D-particles. Using
the results of the previous section they lead to the statistical variances
\beq
\left(\Delta\tilde Y^i_{ab}\right)\left(\Delta\tilde
Y_{ab}^i\right)^\dagger=O_{ab}^{ji}\,O_{ab}^{ki}\,\left(\!\!\left({\cal
Y}_j^{ab}\biggm|\left[{\cal Y}_k^{ab}\right]^\dagger\right)\!\!\right)_{\rm
conn}=\frac{\ell_s\,\Gamma^2}{2\bar g_s^2}\left(1-\frac{\bar
g_s^2}{36}\,\lambda_{ab}^i(u)+{\cal O}\left(\bar g_s^4\right)\right)
\label{variances}\eeq
where the average denotes the connected correlation function with respect to
the probability distribution function ${\cal P}[\rho]$ in \eqn{genussumfinal}.
Note that $\bar g_s^2<0$ owing to the Minkowski signature of the target space
time which is proportional to $\epsilon^{-2}<0$.

The variances \eqn{variances} can now be used to determine a set of uncertainty
relations for the D-particle spacetime coordinates. From
\eqn{eigenaa}--\eqn{Budef} it follows that the resulting uncertainties will
depend non-trivially on the kinematical invariants of the D-brane motion. This
energy dependence is a quantum decoherence effect which can be understood from
a generalization of the Heisenberg microscope whereby we scatter a low-energy
closed string state off the assembly of D-particles. At time $t<0$ we send a
closed string towards the spacetime defects which are fixed in space. As the
closed string hits D0-brane $a$ at $t=0$, it can split into two open strings,
according to the closed-to-open string amplitude formalism of \cite{cardy},
whose other ends can then either attach back to particle $a$ or to D-particle
$b$. Due to this scattering kinetic energy is transfered from the string state
to the D-particles thereby setting them in motion, as depicted in fig.
\ref{stringframe}.

\subsubsection*{Minimum Length Uncertainty}

For a single D-particle, from \eqn{eigenaa} and \eqn{variances} we find the
coordinate smearings
\beq
\Bigl|\Delta{\cal Y}_i^{aa}\Bigr|=|\bar
g_s|^{\chi/2}\,\ell_s\left(1+\mbox{$\frac1{12}$}\,|\bar g_s|^2\,\vec
u_{(a)}^{\,2}\,\delta_{i,1}+\dots\right)\geq|\bar g_s|^{\chi/2}\,\ell_s
\label{smearings}\eeq
where the exponent $\chi\geq0$ is defined through the relation which cancels
the modular divergences of the genus expansion with the tree-level ultraviolet
divergences according to the Fischler-Susskind mechanism \cite{fissuss}
$\log\delta=2|\bar g_s|^\chi\epsilon^{-2}$. It may be fixed upon consideration
of more complicated processes, such as brane exchanges between the D-particles.
It is natural to have $\chi>0$ since the modular divergences are induced by
string interactions. We can use this freedom to fix $\chi$ by the requirement
that the minimum bound in \eqn{smearings} coincides with the 11-dimensional
Planck length $\ell_{\rm P}$, i.e. $\chi=\frac23$. Then the smearings
\eqn{smearings} coincide with standard predictions \cite{liyoneya} based on the
non-relativistic scattering of two D-particles of mass $M_s$ with impact
parameter of order $|\Delta{\cal Y}|$. Note that in the string frame ($i=1$),
the quantum fluctuations exhibit the decoherence effects discussed above.

\subsubsection*{Non-commutative Spacetime Algebra of Observables}

The feature unique to the multi-D-particle system comes from examining
\eqn{variances} for $a\neq b$ which demonstrates explictly the
non-commutativity of the D-particle spacetime. Using \eqn{eigenab} and
\eqn{Oab}, we obtain
from this expression two equations in two unknowns. Outside of the string frame
($i>2$) one finds the same minimal lengths for $\Delta{\cal Y}_i^{ab}$ as in
\eqn{smearings}. Adding these two equations gives smearings $\Delta{\cal
Y}_i^{ab}$, $i=1,2$, in the string frame analogous to \eqn{smearings} that
depend on the center of mass kinetic energy and momentum transfer of the
scattering of D-particles $a$ and $b$ \cite{ms}. Subtracting the two equations
gives an expression for the connected correlation function of the two
coordinate fields of the string plane. This can in turn be written as an
uncertainty relation using the Schwarz inequality
\beq
\Delta{\cal Y}_1^{ab}\,\Delta{\cal Y}_2^{ab}\geq\left|\left(\!\left({\cal
Y}_1^{ab}\Bigm|{\cal Y}_2^{ab}\right)\!\right)_{\rm conn}\right|\geq{\rm
Re}\left(\!\left({\cal Y}_1^{ab}\Bigm|{\cal Y}_2^{ab}\right)\!\right)_{\rm
conn}
\label{schwarzineq}\eeq
where the right-hand side of \eqn{schwarzineq} is found to be
\beq
{\rm Re}\left(\!\left({\cal Y}_1^{ab}\Bigm|{\cal Y}_2^{ab}\right)\!\right)_{\rm
conn}=\frac{|\bar g_s|^{\chi+2}\,\ell_s^2\,\left|\vec u_{(a)}+B(u)\vec
u_{(b)}\right|^2\,{\cal X}_{ab}(u)}{144B(u)\left|\vec
u_{(a)}\right|\sin\theta_{ab}\left(\left|\vec u_{(a)}\right|+B(u)\left|\vec
u_{(b)}\right|\cos\theta_{ab}\right)}~~~~~~,~~~~~~a\neq b
\label{conncorrs}\eeq
Here ${\cal X}_{ab}(u)$ is a complicated function of the D-particle velocities
$\vec u_{(a)}$ and $\vec u_{(b)}$ and their scattering angle $\theta_{ab}$ (see
\cite{ms} for details). The right-hand side of \eqn{conncorrs} vanishes for
zero recoil velocities.

The relation \eqn{conncorrs} gives a non-trivial correlation among different
spatial directions of the target space and represents a new form of
non-commutative spacetime uncertainty relation. It yields the desired
transition from Lie algebraic non-commutativity to a genuine spacetime
non-commutativity, in which the spatial coordinates are no longer independent
random variables due to their string interactions. This is precisely the form
of the description of short-distance spacetime structure based on
non-commutative geometry \cite{ls} which utilizes the algebra of observables of
the quantum string theory. According to \eqn{schwarzineq} the indeterminancies
\eqn{conncorrs} probe much deeper into the exotic short-distance structure than
the usual quantum fluctuation relations. Its energy dependence signifies the
fact that when the D-particles recoil upon impact with a closed string probe
they store information, through the open string degrees of freedom stretched
between them, which prevents independent position measurements for the
D-particles. This leads to correlated spatial uncertainties which depend on the
scattering content, i.e. on the kinetic energies of the non-relativistic
particles. Only when there is no recoil ($\vec u_{(a)}=\vec u_{(b)}=\vec0$) can
one measure simultaneously the positions of two D-particles.

\subsubsection*{Non-commutative Heisenberg Algebra}

We will now describe the quantization of the phase space of the
multi-D-particle system. The canonical quantization condition \eqn{cancomm} on
moduli space leads to the Heisenberg uncertainty principle
\beq
\Delta\bar Y_i^{ab}\,\Delta
P_{cd}^j\geq\mbox{$\frac12$}\,\hbar_\mods\,\delta_i^j\,\delta^a_c\,\delta_d^b
\label{heisenprinc}\eeq
The Planck constant $\hbar_\mods$ can be determined by interpreting
\eqn{genussumfinal} as a minimal uncertainty wavepacket on moduli space
\cite{diffusion} and thereby saturating the lower bound in \eqn{heisenprinc}.
Since the canonical momentum $P_{ab}^i$ is implicitly represented as an
operator on $\mod$ (see \eqn{cancomm}), the effects of the genus expansion on
it are already taken into account and we may therefore compute its variance
directly from the worldsheet $\sigma$-model on a tree-level disc topology.
Using the two-point function \eqn{zammetricexpl} and the one-point function
\eqn{canmompert}, we find
\beq
\left(\Delta
P_{ab}^i\right)^2=G_{ab;ab}^{ii}-\left(P_{ab}^i\right)^2=\frac{4\bar
g_s^2}{\ell_s^2}\,\delta_{ab}+\frac{2\bar
g_s^2}{9\ell_s^2}\left(2\delta_{ab}\left[\left(\bar
U^i\right)^2\right]_{ba}-287\left(\bar U^i_{ba}\right)^2\right)+\dots
\label{momvariance}\eeq
Performing a Galilean boost to a comoving target space frame, i.e. setting
$\bar U^i=0$ in \eqn{momvariance}, and using the minimum length \eqn{smearings}
in \eqn{heisenprinc} then determines the Planck constant as
\beq
\hbar_\mods=4\,|\bar g_s|^{1+\chi/2}
\label{hbar}\eeq
Thus, the Planck constant in the present formalism is proportional to the
string coupling constant, which owes to the fact that the quantization of
$\mod$ here is induced by string interactions.

Note that to leading order the operator \eqn{canmompert} coincides with the
spacetime momentum $p^i=M_s\bar U^i$. However, stringy effects give corrections
to this operator of the form $P^i=p^i+{\cal O}(\bar g_s^2)(p^i)^3$. Iterating
the Heisenberg algebra \eqn{cancomm} using this identification along with
\eqn{canmompert} gives the string modified phase space commutation relations
\cite{ms}
\bea
\left[\!\left[\widehat{Y}_i^{ab},\widehat{p}^{\,j}_{cd}\right]\!\right]&=&i\,
\hbar_\mods\left(\delta_i^j\delta_c^a\delta_d^b+\mbox{$\frac1{96}$}\,|\bar
g_s|^2\bar\ell_s^{\,2}\left[\delta_i^j\left(\delta^a_c\,[\widehat{p}_k^{\,2}
]_d^b+\delta_d^b\,[\widehat{p}_k^{\,2}]^a_c+[
\widehat{p}_k]^a_c[\widehat{p}^{\,k}]_d^b\right)\right.\right.\nn\\& &
\left.\left.+\,\delta^a_c\left\{\widehat{p}_i,\widehat{p}^{\,j}
\right\}^b_d+\delta_d^b\left\{\widehat{p}_i,\widehat{p}^{\,j}\right\}^a_c+
[\widehat{p}_i]^b_d[\widehat{p}^{\,j}]^a_c+[\widehat{p}_i]^a_c
[\widehat{p}^{\,j}]^b_d\right]+\dots\right)
\label{strmodheisencomm}\eea
to leading orders, where $\bar\ell_s=\ell_s/|\bar g_s|^2$ is the (time
independent) 0-brane scale. For $a=b=c=d$ and $i=j$, \eqn{strmodheisencomm}
yields the standard string-modified Heisenberg uncertainty principle
\eqn{stringheisen} \cite{ven} for a single recoiling D-particle \cite{lm,kmw}.
However, for the off-diagonal degrees of freedom, \eqn{strmodheisencomm} takes
into account of the string interactions among the D-particles and represents
the phase space version of the non-commutative correlators that we obtained
above. It would be interesting to study the representation theory of the
algebra \eqn{strmodheisencomm} and thereby determine properties of the
non-commutative spacetime algebra of observables implied by the relations
\eqn{schwarzineq} and \eqn{conncorrs}.

\subsubsection*{Space--time Uncertainty Principle}

The target space time $T$ in the ``physical'' frame is given by \eqn{phystime}.
Upon summing over all worldsheet genera, the promotion of the couplings $\bar
Y$ and $\bar U$ to operators implies that $T$ becomes an operator $\widehat T$.
To leading orders in the string coupling constant expansion, we may replace the
velocity operators in \eqn{calUdef} by momentum operators according to
\eqn{canmompert}, as described above. Using \eqn{cancomm} and the present
Born-Oppenheimer approximation to expand the function \eqn{calUdef} as a power
series in $\bar U_{ab}/u_c\ll1$, $a\neq b$, we arrive at the space--time
quantum commutators \cite{ms}
\beq
\left[\!\left[\widehat{Y}_i^{ab},\widehat{T}\right]\!\right]=\frac{i\,
\ell_s^2\,\hbar_\mods}{2|\bar g_s|}\left[\delta^{ab}+\left(1-\delta^{ab}
\right)\frac{\ell_s}{4|\bar g_s|}\frac{\widehat{P}_j^{ab}}{\sqrt E}+\dots
\right]
\label{YTcomm}\eeq
to leading orders, where $E=|\bar g_s|^2\sum_a(\bar U_{aa}^i)^2$ is the total
kinetic energy of the constituent D-particles. Using \eqn{hbar} we thus obtain
the space--time uncertainty principle
\beq
\Delta\bar Y_i^{aa}\,\Delta T\geq|\bar g_s|^{\chi/2}\,\ell_s^2
\label{spacetimeuncert}\eeq
When $\chi=0$ the indeterminancy relation \eqn{spacetimeuncert} coincides with
the standard one \cite{liyoneya} which can be derived from the energy-time
uncertainty principle of quantum mechanics applied to strings. It also follows
from very basic worldsheet conformal symmetry arguments and it gives a natural
representation of the $s$-$t$ duality of perturbative string amplitudes. In the
present case this uncertainty relation follows directly from the phase space
uncertainty principle and it shows that there is a duality between short and
large distance phenomena in string theory. However, the choice $\chi=0$ gives a
minimum length \eqn{smearings} which is much larger in general than the
11-dimensional Planck scale. The ambiguities here follow from the fact that the
physical target space (Liouville) time coordinate $T$ is not the same as the
longitudinal worldline coordinate of a D-particle, but is rather a collective
time coordinate of the system of particles which is induced by all of the
string interactions among them. We can nevertheless match our results with
those of 11-dimensional supergravity by multiplying the definition
\eqn{phystime} by an overall factor of $|\bar g_s|^{-\chi/2}$, which then
implies that the target space propagation time for weakly-interacting
D-particles is very long.

\subsubsection*{Triple Uncertainty Relations}

The commutation relation \eqn{YTcomm} for $a\neq b$ illustrates the effects of
the string interactions on the space--time duality relation. Using the
canonical minimal uncertainty \eqn{heisenprinc} and rescaling the time
coordinate $T$ as described above, we arrive at the triple uncertainty
relations
\beq
\left(\Delta\bar Y_i^{ab}\right)^2\,\Delta T\geq\frac{|\bar
g_s|^\chi\,\ell_s^3}{2\sqrt E}~~~~~~,~~~~~~a\neq b
\label{tripuncert}\eeq
This uncertainty principle implies that the high-energy scattering of
D-particles can probe distances much smaller than the characteristic length
scale in \eqn{tripuncert}, which for $\chi=\frac23$ is $\ell_{\rm P}\ell_s^2$.
Triple uncertainty relations of the sort \eqn{tripuncert} but involving only
$\ell_{\rm P}^3$ have been suggested based on the holographic principle of
$M$-theory \cite{liyoneya}. The existence of a limiting velocity $|\vec u_a|<1$
for the non-relativistic D-particle motion implies a lower bound on
\eqn{tripuncert}, so that using the minimum spatial extensions \eqn{smearings}
and setting $\chi=\frac23$ in \eqn{tripuncert}, we arrive at the characteristic
temporal length
\beq
\Delta T\geq|\bar g_s|^{-1/3}\,\ell_s
\label{templength}\eeq
which also agrees with the standard result based on D-particle kinematics
\cite{liyoneya}.

\newsection{Outlook: Potential Experimental Tests of Spacetime
Non-commutativity}

We have seen that a perturbative worldsheet formalism for systems of D-branes
yields results which are consistent with the standard target space dynamics,
and which also describes interesting new short-distance structures, such as
non-commutative spatial coordinates that lead to a proper spacetime
quantization. The most dramatic feature of the uncertainty relations we
exhibited in the previous section is their dependence on the energy of the
D-particle system. This fact distinguishes D-particle dynamics from ordinary
quantum mechanics, since it implies a bound on the accuracy of length
measurements which depends entirely on the energy content of the system. Such a
situation, whereby the accuracy with which one can measure a quantity depends
on its size, is a characteristic feature of decoherence in certain approaches
to quantum gravity. In the present case the presence of D-brane domain wall
structures may act as traps of low energy string states, thereby resulting in a
decoherent medium of quantum gravity spacetime foam. The quantum coordinate
fluctuations, due to the open string excitations between D-particles, can lead
to quantum decoherence for a low-energy observer who cannot detect such recoil
fluctuations in the sub-Planckian spacetime structure. The short-distance
physics described by non-abelian D-particle dynamics in flat target spaces
naturally capture features of spacetime quantum gravity, and the construction
outlined above therefore illuminates the manner in which D-particle
interactions probe very short distances where the quantum nature of gravity
becomes important.

It would be interesting to see if the non-commutativity of quantum spacetime is
amenable in some way to experimental analysis. The foamy properties of the
non-commutative structure may require a reformulation of the phenomenological
analyses of length measurements as probes of quantum gravity. One such test is
through neutral kaon systems \cite{dafne} which are sensitive to the minimal
length suppression effects by the Planck mass scale $M_{\rm P}\sim10^{19}$ GeV,
and also to quantum gravity decoherence effects. A more recent suggestion is
through $\gamma$-ray burst spectroscopy \cite{grb}. Such probes are
cosmological in origin and are sensitive to Planck scale energies through
quantum gravity dispersion relations in which the velocity of light depends on
the photon energy. However, all of these approaches do not incorporate length
measurements in the transverse directions to the probe, so that it is not
immediately clear how to incorporate non-commutative correlations such as
\eqn{conncorrs} into these analyses.

A recent proposal \cite{percival} which is intimately related to the ideas of
this article is that fluctuations in spacetime geometry on the scale of the
Planck time $\tau_{\rm P}\sim10^{-44}$ s may be detectable by atom
interferometers. In analogy with Brownian motion, whereby measurements on a
macroscopic scale can be used to determine quantities on an atomic scale, one
can find a diffusion process which enables the determination of quantities at
the Planck scale by experiments at an atomic scale. Spacetime fluctuations
induce diffusion in quantum amplitudes from which the value of $\tau_{\rm P}$
can be measured and information about Planck scale dynamics can be extracted.
The key feature of this analysis is an appropriate generalization of linear
Markovian quantum state diffusion to non-commuting fluctuation variables which
span an isospin space of internal symmetries of the spacetime that is distinct
from the ordinary position space. The canonical commuting fluctuations yield no
effect in matter interferometers, but the decoherence effects resulting from
non-commutative fluctuations lead to a suppression of the observed
interference. The analysis of \cite{percival} thus shows that the small
numerical value of the Planck time does not on its own prevent experimental
access to Planck scale physics in the laboratory. The resulting non-commutative
metric is augmented into the isospin space which is attached to the original
spacetime itself and not to the matter within it. Thus if we consider
D-particles as being intrinsic topological defects of spacetime representing
short-distance singularities of quantum gravity, then the non-commutativity
described in this paper may be related to the description of \cite{percival}.
In \cite{diffusion} the relationship between D-brane recoil and diffusion in
open quantum systems is discussed. It would be interesting to explore the
potential relationship with D-particle spacetimes and those of \cite{percival}
in more
detail, and hence establish an experimental laboratory for the Planck scale
dynamics probed by D-branes.

\end{document}